Polarization dependence of coherent phonon generation and detection in the 3D topological insulator $Bi_2Te_3$

O.V. Misochko[1], J. Flock[2], and T. Dekorsy[2]

[1] Institute of Solid State Physics, Russian Academy of Sciences

142432 Chernogolovka, Moscow region, Russia

[2] Department of Physics, University of Konstanz, Universitatsstrasse 10, D-78457 Konstanz, Germany

We have studied the polarization dependence of coherent phonons in the topological insulator $Bi_2Te_3$. Using polarization-dependent femtosecond pump-probe spectroscopy, we measured coherent phonons as a function of angle when the pump and probe polarizations were fixed, and the crystal orientation was rotated. For isotropic detection, depending on the spot position, oscillations either from only low- and high-frequency phonons of $A_{1g}$ symmetry, or in addition from the mode at 3.6 THz were observed. All the modes were found to be independent of the orientation of electric field vector with respect to the crystal axes testifying to their full symmetry while no modes of lower symmetry appeared in any polarization geometry. For anisotropic detection both modes of $E_g$ symmetry could be detected, but their amplitudes were considerably smaller than those of $A_{1g}$ symmetry. To clarify the coherent phonon assignment and the process of coherent phonon generation in $Bi_2Te_3$, the time-domain measurements were complemented by spontaneous Raman scattering. The comparison of frequency- and time-domain results and the polarization dependence suggest that the 3.6 THz mode belongs to crystalline Te arising due to tellurium segregation. A discrepancy between the time- and frequency domain data is discussed.

I. Introduction

Topological insulators are new materials with surfaces that host a new state of matter insensitive to impurities and defects [1]. In these materials surface electrons behave like massless Dirac particles and surface currents preserve their spin orientation and coherence on a macroscopic scale. The topological insulators were suggested to be used in quantum computing because they contain surface states that are topologically protected against scattering by time-reversal symmetry. Moreover, they were also proposed for applications in memory devices where write and read operations can be achieved by purely electric means. The growing interest in the topological materials resides not only in their promising properties for future "topological" applications, but also in basic research. For instance, their fundamental properties apart from such exotic particle as Dirac fermion suggest a relation to elusive Majorana fermion [2] that is its own antiparticle.

Optical techniques are reliable tools for the study of electronic and lattice properties of solids. Indeed, optical spectroscopy has obtained valuable information on the charge dynamics of the Dirac fermions in graphene [3], and several works [4-6] have predicted interesting interactions between topological insulators and light. Electron-phonon interaction can be a dominant scattering mechanism for the Dirac fermions at finite temperatures as technical improvements may minimize crystal imperfections, but phonons are always present and therefore limit the feasibility of "topological" applications. Phonons in solids can be measured via Raman scattering and infrared spectroscopy, which measure the lattice modes at the Γ-point of the Brillouin zone. These techniques can be extended into the time domain, but their time resolution is typically low. The frequency-domain techniques providing the valuable information on thermal phonon frequencies and their lifetimes are inherently unable to resolve individual cycles of atomic vibrational motion even though they do allow the determination of the symmetry of different lattice modes. To overcome this limitation, coherent phonon generation by femtosecond laser pulses can be employed. Such femtosecond time-domain techniques in which a ultrashort pump pulse generates coherent oscillations that modify the dielectric function, and a probe pulse samples the crystal at a later time, can be used to generate, control and detect coherent phonons for which vibrational dynamics are time resolved to a fraction of a single vibrational period [7]. Instead of analyzing frequency domain line shifts and line shapes, where all information on the incoherent lattice dynamics is hidden, one can nowadays visualize atomic motion and observe both the atomic oscillations and their decays in real time. Moreover, these synchronized atomic motions due to the existence of well-defined phase are suitable for coherent control that can allow manipulating atomic motions and electron-phonon interaction in future topological devices.

Bismuth telluride $Bi_2Te_3$ has been classified as one of 3D topological insulators characterized by a single Dirac cone [8]. Its lattice dynamics was thoroughly studied in the frequency domain [9-11], and there were a number of recent time-domain studies of both carrier and lattice dynamics [12-16]. Nevertheless, several aspects of coherent lattice dynamics in $Bi_2Te_3$ still remain controversial. For example, in Ref. [12], the two observed coherent phonon modes at 1.85 THz and 3.68 THz were assigned as $A_{1g}^{(I)}$ mode and its second harmonic. In later studies [13, 14], the modes at 1.85 THz and 4.02 THz were assigned to two allowed $A_{1g}$ modes, in agreement with spontaneous Raman measurements [9,10]. The detailed time-domain study [15] in which both the temperature and power dependence of coherent phonons were measured confirmed that the higher-frequency coherent phonon mode, which appears with 3.6 THz frequency at room temperature and increases to 4.0 THz at helium temperature, is an $A_{1g}^{(II)}$ mode and not a second harmonic of the $A_{1g}^{(I)}$ mode. This high frequency mode acquires two orders of magnitude higher positive chirp at the lowest temperatures and high pump fluence as compared to the $A_{1g}^{(I)}$ phonon. Our previous study [16] revealed that depending on the pump polarization with respect to the crystal orientation one can see either only two fully symmetric coherent phonons at 1.8 and 4.0 THz, or, in addition, the mode at 3.6 THz assigned tentatively to a doubly degenerate $E_g^{(II)}$ phonon. However, the symmetry of the 3.6 THz mode was not proved as it requires the full knowledge of its angle dependence missing in [16].

The objective of the present work is to study coherent phonons in $Bi_2Te_3$ crystal as a function of angle between the crystal axis and the beam polarization. It is aimed to clarify the assignment of the 3.6 THz mode and, additionally, by comparison of frequency- and time-domain results to elucidate the mechanism of coherent phonon generation in topological insulators. The paper is organized as follows: In sections II and III we address the technical aspects regarding the setups for Raman spectroscopy and pump-probe technique and the properties of $Bi_2Te_3$ crystals, which are related to coherent lattice dynamics. In section IV we show and analyze our experimental results obtained first in the time-domain and then in the frequency-domain. Finally, section V contains the summary of the results, and the conclusions.

II. Experimental

In this study we used the same rectangular piece of $Bi_2Te_3$ crystal as in [16] with a surface perpendicular to the trigonal axis. Mechanical exfoliation was performed with an adhesive tape to obtain a surface with good optical quality. The zero polarization angle corresponds to the electrical vector of the laser light oriented along the short dimension of the crystal, which is at 15 degrees to the binary axis. The time-domain data were measured in a pump-probe setup based on asynchronous optical sampling [17,18]. It consisted of two Ti:sapphire oscillators with repetition rates in the GHz range and a kHz offset frequency in the repetition rate. Each oscillator provided 50 fs pulses with wavelength of 818 nm for the pump and of 830 nm for the probe. The crystal was excited with 50 mW average pump power focused to a spot with a diameter of 60 µm (typical pump fluence 1.6 µJ/cm$^2$, much less than the damage threshold of 4.5 mJ/cm$^2$ [15]). The changes in reflection intensity were sampled with 200 MS/s (megasamples per second) on a detector with a bandwidth of 125 MHz. In most experiments we used isotropic detection scheme in which the differential reflection of the probe, $\frac{\Delta R}{R_0} = \frac{R - R_0}{R_0}$ was defined as the relative change in the reflection ($R$) caused by the pump pulse, where $R_0$ is the reflection at negative time delays without the pump pulse. In a few experiments, we employed anisotropic detection scheme in which the reflected probe beam was analyzed into polarization components parallel $R_\parallel$ and perpendicular $R_\perp$ to that of the pump, and their difference ($R_\parallel - R_\perp$) was recorded with two photodiodes. Transient reflectivity was detected with 10 mW probe light focused to a spot with 20 µm in diameter. Each transient was the average over 2 million traces taken over a time span of 17 minutes. To study the polarization dependence of the ultrafast response we rotated the sample about its surface normal, while the pump and probe light polarizations remained fixed and parallel to each other for isotropic detection. In some measurements in order to keep the excitation spot unchanged we used λ/2 plates to rotate the pump and probe polarizations while the crystal orientation remained fixed. The scanned time interval was larger than 300 ps in all the experiments. In a few experiments, the transient reflectivity initiated by 800 nm light was measured with a mechanical delay stage in a shorter time interval using the similar isotropic detection scheme.

Raman spectra were recorded on a micro-Raman spectrometer (Microdil-28) used in a backscattering configuration. The spectra were excited with a visible laser light ($\lambda$= 632.8 nm) at low power levels, typically less than 1 mW, measured with the power meter at the sample position. The low power levels were essential to avoid local laser heating and damage, which were much more pronounced than in other crystals due to the extremely low thermal conductivity of $Bi_2Te_3$. In a few experiments the spectra were taken with second harmonic of Nd:YAG laser ($\lambda$ = 532.1 nm). All the spectra were collected through a 50×objective and recorded with 1800 lines/mm grating and slit widths providing a spectral resolution better than 1 cm$^{-1}$. The incident laser excitation was linearly polarized and travelled along a path that includes a half-wave plate for rotating the plane of polarization, the microscope beamsplitter, and the objective. The spectra were obtained at room temperature by using the multichannel, nitrogen-cooled CCD detector situated after the analyzer and the spectrograph. They were typically acquired after 100 accumulations with 10 s integration time.

III. Bismuth telluride properties related to coherent phonons

$Bi_2Te_3$ belongs to a class of the simplest 3D topological insulators whose surface states consist of a single Dirac cone at the $\Gamma$- point [1]. The surface states are formed by $p_z$ atomic orbitals, and the neighboring bulk conduction band and bulk valance band, separated by 0.21 eV gap, are formed by $p_x$ and $p_y$ orbitals. Strong spin-orbit interaction locks the spin of the Dirac fermion and its wave-vector in mutually perpendicular directions, giving the Dirac cone a definite chirality. Bismuth telluride has a rhombohedral unit cell belonging to the space group $D_{3d}^5(R\bar{3}m)$. The primitive unit cell contains one formula unit of five atoms. Its binary axis with twofold rotation symmetry is along $x$, while the bisectrix axis occurring in the reflection plane is along *y*. Usually, the description of $Bi_2Te_3$ is based on the equivalent hexagonal unit cell, obtained by choosing the *z* axis in the [111] rhombohedral direction. In this case, the crystal exhibits a layered structure with stacking of the sequence $Te^1$-Bi-$Te^2$-Bi-$Te^1$, where the superscripts label two different positions of Te atoms in a quintuple layer. Bonding between atomic planes within a quintuple layer is covalent, whereas bonding between adjacent layers is predominantly of the Van der Waals type. The $Te^2$ atom lies at an inversion center, which possesses $D_{3d}$ symmetry (1(a) Wyckoff position), while $Te^1$ and Bi atoms occupy $C_{3v}$ sites (2(c) Wyckoff positions). As the primitive unit cell contains five atoms, there are 15 normal phonon modes. At the $\Gamma$-point they are classified according the group theory [11] to $2(A_{1g}+E_g)+3(E_u+A_{2u})$. Two $E_u$ and one $A_{2u}$ are acoustic, transverse and longitudinal, modes, respectively, and the rest are optical ones. Since the symmetry group possesses an inversion center, exclusion rule applies. As a result, the optic modes are either Raman (*gerade*) or IR (*ungerade*) active and both the Raman and IR spectra consist of four peaks each. In the doubly degenerate $E_g$ modes, the atoms vibrate in the basal *xy* plane, while in the both $A_{1g}$ modes the atoms move along the trigonal direction *z*. Raman tensors of the modes are as follow:

$$A_{1g} = \begin{bmatrix} a & & \\ & a & \\ & & b \end{bmatrix}; \quad E_g = \begin{bmatrix} & -c & -d \\ -c & & \\ -d & & \end{bmatrix} \text{ and } \begin{bmatrix} c & & \\ & -c & d \\ & d & \end{bmatrix},$$

from which one can see the *xy*- or the $x^2-y^2$ symmetries in the basal plane for doubly degenerate modes. The difference between the $E_g^{(I)}$ and $E_g^{(II)}$ modes is that the top two layers vibrate in phase or out of phase and the frequency of the in-phase vibration is low, and the out-of-phase is high [16]. The similar behavior is also true for the two fully symmetric modes as schematically shown in Fig.1

IV. Results and discussion.

We started our study with the measurements in the time-domain. Figure 2 a shows the typical experimentally observed transient reflectivity $\Delta R/R_0$ of $Bi_2Te_3$ crystal at room temperature. The signal consists of a nonoscillatory background (the initial fast drop, and a slower recovery, both related to electron excitation and/or lattice heating via electron-lattice coupling) and oscillatory components appearing right after laser excitation. A fast Fourier Transform (FT) of the data reveals fast and slow oscillations at 4.04 THz and 1.84 THz, corresponding to the frequencies of $A_g^{(II)}$ and $A_g^{(I)}$ phonon modes. The frequencies are in agreement with previously reported Raman experiments on $Bi_2Te_3$ single crystals [9,10] and the calculations based either on Born-von Karman lattice model [19], or on *first-principles* calculations [20].

We first fit the transient reflectivity in the time-domain. To this end we differentiate the signal to get rid from the incoherent component and fit the derivative in the range from zero delay to 15 ps to two damped harmonic oscillators, see Fig. 3. The decay rates and frequencies obtained are listed in Table 1. If we restrict the range of fitting to 3 ps, which contains only several oscillations, the frequencies for both modes exhibit a small blue shift. This shift is indicative of phonon chirp, the situation when the frequency for small delays is larger than the frequency for long delays. Line shapes of the fully symmetric modes in FT spectrum shown in Fig.2 b are asymmetric also signaling that the phonons are chirped (i.e., phonon frequency is a function of time). Such line shapes are quite similar to the Fano profiles, which can be deduced from the time-domain data when a discrete phonon state couples to an electronic continuum [21,22]. The fits to the Fano profiles $\sigma(\varepsilon) \propto \frac{(q+\varepsilon)^2}{1+\varepsilon^2}$, where ε is a dimensionless energy normalized to decay $\Gamma$, and $q$ is the asymmetry parameter, are shown in Fig.2(b) by solid lines. The frequency domain is a better choice for identifying the chirp as it is difficult to see the chirp in the time-domain, especially for the high frequency mode, which is short lived [15]. From the FT spectra one can see that both fully symmetric phonons are negatively chirped as low frequency slope of the lines is steeper (Fano parameter is positive). The positive value of $1/q$ implies that there is a constructive interference between the discrete phonon and electronic continuum on the high-energy side and a destructive interference on the low-energy side. Nevertheless we do not know at the moment which continua participate in this coupling to the $A_{1g}$ modes. From their asymmetry quantified by the inverse of the Fano parameter $1/q$ follows that the chirp is a bit larger for low frequency mode. However, the lifetime of the negative chirp is very short for each mode. This short lifetime can be inferred from the fact that the line shape asymmetry (and hence the phonon chirp) completely disappears after one or two oscillation cycles for both modes as it is shown by a few FT spectra obtained with the left edge of time window for FT shifted, see Fig.4. Fitting the transient reflectivity data in the time-domain by two damped harmonic oscillators we identified that the low frequency mode had a larger initial amplitude and a smaller decay rate, see Table 1. The low frequency mode had a larger initial amplitude and a smaller decay rate as can be seen from the FT spectra shown in Fig. 2(a). The factor of 4 difference between the amplitudes of the low- and high frequency modes arises because the Raman polarizability for the low frequency mode is larger and its frequency is smaller than those for high frequency mode. The latter fact results in more efficient generation arising due to exponential dependence of the initial amplitude $A$ on pulse duration $A \propto e^{-\frac{\Omega^2 \tau_p^2}{4}}$ [23], where $\Omega$ is the phonon frequency and $\tau_p$ is the laser pulse duration. The ratios of decay rate to mode frequency, i.e. Fano factor [24] for both modes, are almost the same. Note that a similar observation has been reported for coherent phonons in $Bi_2Se_3$ [25, 26], however, initial amplitudes of the coherent modes obtained by anisotropic detection, i.e. the subtraction of two orthogonal components of the reflected linear polarized light, were reported to be almost equal.

Further measurements were made as the planes of polarization for the pump and probe beams were varied relative to the crystallographic axes of the sample. As the crystal was rotated, no variation of frequency, amplitude and phase of the coherent oscillations was observed demonstrating the lack of polarization anisotropy for the both modes. Results of the polarization-dependent ultrafast pump-probe measurements are shown in Fig. 5 where one can see that the amplitude ratio of the modes is constant, independent of the polarization angle. Moreover, we should emphasize that at any polarization angle no modes of $E_g$ symmetry were detected. This is natural for the low symmetry ($xy$) mode, which cannot be excited when the pump and probe polarizations are parallel and therefore cannot couple to the phonons whose tensor does not have a diagonal component. However, this is not true for the $E_g$ symmetry mode with the $x^2 - y^2$ symmetry whose tensor has a zero trace due to different signs of the diagonal components. In this case, the two degenerate $E_g$ phonon modes rotate the polarization of the probe pulse by the same amount in opposite directions, giving rise to a zero net modulation of the probe pulse's polarization only when the light polarization is at π/4 respective to binary, or bisectrix axes. The lack, or at least extremely small contribution of $E_g$ modes to the transient reflectivity, suggests that there are some peculiarities in the generation mechanism of the coherent, low-symmetry modes.

Compared to our previous study of the same crystal [16] we did not observe the 3.6 THz mode at any angle. This mode had been tentatively assigned to $E_g^{(II)}$ mode and thought to appear in the transient reflectivity due to its predominant $x^2 - y^2$ symmetry [16]. Therefore, we scanned the crystal surface to find the regions where the mode to be present. Scanning the crystal surface we observed that for certain locations the transient reflectivity changed considerably. Figure 6(a) shows a representative transient reflectivity obtained for these locations. The fast negative drop at zero delay for these locations is twice smaller than in Fig.2(a), while its recovery time is shorter. FT spectrum of these regions, apart from 2 intrinsic $Bi_2Te_3$ phonon modes with slightly reduced amplitudes, has an additional strong line at 3.56 THz, see Fig.6 (b). It should be stressed that in our previous study [16], the observation of the 3.6 THz mode was also accompanied by a reduced negative drop at zero delay. In contrast to the data shown in Fig.2(b), the fully symmetric phonons as well as the 3.56 THz mode seem to be unchirped because their line shapes are symmetric. At the same time, the differences in the frequencies observed from the different spot locations were negligibly small suggesting the presence of similar bonds.

Rotating the crystal, we observed that the amplitude of the 3.56 THz mode was independent of the angle, thus testifying to its full symmetry, see Fig.5. Finding the angle independence excludes the possibility of the 3.56 THz mode assignment to a low symmetry phonon made in [16]. Here we would like to emphasize that at any polarization angle no modes of $E_g$ symmetry were observed for those regions where the 3.56 THz mode was present. Thus combining this observation with the previous results shown in Fig.2, we established that the phonons of low symmetry could not be coherently excited and detected in the present configuration, or their amplitudes were over an order of magnitude smaller than those of fully symmetric phonons.

We note that similar 3.6 THz mode has been previously observed in a number of time-domain studies on $Bi_2Te_3$ [12,15,16]. For instance, in [12,15] this mode appeared at high fluence excitation and, moreover, its frequency was downshifted by increasing the excitation power. Such behavior reminds of fully symmetric mode of Te, which is known to decrease in frequency due to electronic softening [27,28]. The decrease resulting in a positive phonon chirp is linear proportional to the laser fluence [29]. Thus, it is reasonable to suggest that this mode appears in $Bi_2Te_3$ materials due to tellurium segregation in Te-rich regions. To check this hypothesis we carried out the time-domain study of single crystal of Te. Its crystal structure consists of three-atom per turn helices whose axes are arranged on a hexagonal lattice. Group theory predicts four zone centre optical phonons of which three are Raman active. These are two doubly degenerate E modes (at 2.7 and 4.2 THz) and one non-degenerate $A_1$ mode. The fully symmetric $A_1$ phonon at 3.57 THz is a "breathing" mode of the Te lattice for which the helical radius changes leaving the interhelical distance and $c$-axis spacing intact, thereby preserving the lattice symmetry. We coherently excited $A_1$ phonons in Te single crystal with a pump polarized perpendicular to the trigonal axis. As one can see from Fig.7, the transient reflectivity of Te consists of two contributions. The oscillations due to coherent phonons are superimposed on an exponential background arising from electronic excitations. Note that the polarity of the electronic contribution in Te is positive, in a contrast to that of $Bi_2Te_3$. The strong modulation of the reflectivity occurs through coherent $A_1$ oscillations with a frequency of 3.57 THz, as shown in Fig.7(b).

Having finished with the time-domain measurements, we now discuss the study of lattice dynamics for $Bi_2Te_3$ in the frequency-domain in order to get additional confirmation to the phonon assignment by comparing the time- and frequency-domain data. First, we measured the regions where only two fully symmetric modes had been observed in the time-domain. The spectra of Fig. 8 show Raman data with two $E_g$ and two $A_{1g}$ phonons for $\lambda$ = 632.8 nm excitation at room temperature. The peak intensity of $A_{ug}^{(II)}$ mode is comparable to the intensity of $E_g^{(II)}$ mode and twice smaller than that of $A_{ug}^{(I)}$ mode. The phonon intensities varied moderately when the excitation line was changed to $\lambda = 532.1$ nm due to resonance effects, however, the phonon lines remained sharp and the spectra showed no background at any excitation wavelength. The upper spectrum in Fig. 8 was recorded with parallel polarizations of the incoming and scattered light (VV configuration) both coinciding with the short dimension of the crystal. The doubly degenerate $E_g$ modes can be easily distinguished from fully symmetric ones due to zero trace of the Raman tensor for the modes that are projected in measurements for crossed (VH configuration) polarizations of incident and scattered light (depolarized spectrum). The $E_g^{(I)}$ mode is the weakest one with intensity a factor of 4 smaller than that of $E_g^{(II)}$ mode. The $A_{1g}^{(I)}$ and $E_g^{(II)}$ modes have narrower line widths as compared to the $A_{1g}^{(II)}$ x and $E_g^{(I)}$ modes. In Table 1 we present frequencies and decay rates obtained by fitting the phonon lines to Lorentz profiles.

Scanning the crystal surface we observed that the Raman spectra for certain locations were significantly different from those where only intrinsic phonons were detected. For the former locations, apart from 4 intrinsic phonon modes of $Bi_2Te_3$, an additional strong line at 3.6 THz and two weaker lines at 2.7 and 4.25 THz as well as some electronic background are clearly seen in Fig.9. Moreover, the relative intensities of intrinsic phonons for these spot locations are slightly modified. To assign the additional lines we measured Raman spectrum of (0001) Te single crystal shown in Fig.9 by a solid line. The comparison between the two spectra suggests that the extra lines observed in some regions belong to crystalline tellurium. Their similar (to tellurium) line widths testify to a high crystallinity of theseTe-rich regions, whereas different relative intensities of the fully symmetric $A_1$ and doubly degenerate E phonons can be due to unknown microcrystal orientation. Thus, we tentatively ascribe the appearance of the additional lines in $Bi_2Te_3$ to Te segregation. The realization of Te-rich composition, with respect to stoichiometric $Bi_2Te_3$, remains unknown. Note that similar tellurium segregation has been reported for Bi-Te films [30] and for $Sb_2T_3$ films under strong laser irradiation [31]. Previously, the two vibrational modes at 2.8 and 3.67 THz were observed in Raman scattering for $Bi_2Te_3$ nanoplates, but not in bulk crystals [32]. They were ascribed to IR modes and their appearance was interpreted as evidence of a parity breakdown in thin flakes of topological insulator.

To clarify the origin of Te-rich regions and their relation to the exposure of laser light we intentionally damaged the crystal by intense laser light. Close to the burned area we observed in a microscope a lot of black cuboid nicrocrystals scattered over the crystal surface. The density of the microcrystals decreased as the distance from the damaged area increased. The time-domain measurements carried out near the damaged area shown in Fig.10 revealed that the negative drop of initial reflectivity changed the polarity and the FT spectrum demonstrated the dominant contribution from the 3.56 THz mode. This peak rapidly became the unique features present in the spectrum for spot locations closer to the burned area. Therefore, we can speculate that since Te atoms are lighter and more volatile than Bi atoms, they become mobile during excitation at high laser powers. The dominant behavior of the 3.56 THz mode is to be ascribed to a greater Raman polarizability for Te than for Bi or $Bi_2Te_3$. Thus, it is thought that the 3.56 THz mode in the time-domain data of $Bi_2Te_3$ to be the $A_1$ mode of crystalline Te that forms through a segregation of tellurium.

Finally, let us address the issue why no low symmetry mode appears in our time-domain data even though their Raman polarizabilities are almost the same as those for fully symmetric modes as evidenced by the frequency-domain study. We have to stress here that the generation of low symmetry coherent phonons in $Bi_2Te_3$ is suppressed but not prohibited. Indeed, it is appropriate to mention that in $Bi_2Se_3$ and $Sb_2Te_3$ topological insulators, which are isomorphic to $Bi_2Te_3$, both $E_g^{(I)}$ and $E_g^{(II)}$ modes have been detected [25,26,33] using anisotropic detection scheme. To access the anisotropic reflectivity change, after reflecting from the sample, the probe beam is analyzed into polarization components parallel and perpendicular to that of the pump, $R_\parallel$ and $R_\perp$, and their difference is recorded with two balanced photodiodes. This scheme is especially useful in isolating small anisotropic signals from large isotropic ones as vertically and horizontally polarized parts of the probe beam see different modulations induced by the anisotropic excitation but the same modulations induced by the isotropic excitation. As a result, no fully symmetric mode with equal diagonal elements can be detected by anisotropic detection; however, the amplitude of fully symmetric phonons in [26] was comparable with that of $E_g^{(II)}$ mode. This can be due to either unbalanced photodiodes, or coupling to the "$b$" component of the $A_{1g}$ Raman tensor and excludes the possibility to compare polarizabilities of the modes correctly from the anisotropic transient reflectivity alone. Recall that in our time-domain study we used isotropic detection that samples the diagonal components of Raman phonons and therefore should allow the comparison of "$a$" and "$c$" components of Raman tensors from a single trace. Nevertheless, to make contact with other topological insulators we decided to check the generation of coherent phonons in $Bi_2Te_3$ with anisotropic detection. To this end, a half waveplate was inserted in the probe beam to change its polarization to $\pi/4$, while the polarization of the pump beam was kept horizontal. The reflected probe beam was split by a polarizing cube beamsplitter into two beams with vertical and horizontal polarizations, and then sent to the two detectors. One of the detectors recorded the signal analogous to that in isotropic detection, see Fig.11(a), whereas the difference signal shown by the lower trace in Fig.11(a) revealed low symmetry oscillations. As shown in Fig.11(b), where normalized FT spectra are presented, the oscillations of the difference signal contain frequency close to 3 THz. In addition, there is a very weak peak around 1 THz, which is just close to noise level. Both frequencies coincide with those of $E_g$ modes. One order of magnitude difference between their amplitudes cannot be due to different Raman polarizabilities since the exponential dependence of the initial amplitude on the ratio of pulse duration to phonon period for the low frequency $E_g^{(I)}$ mode should makes their amplitude almost equal given the factor of 3 larger frequency for the $E_g^{(II)}$ mode. Furthermore, their amplitudes, in contrast to those of $A_{1g}$ modes, demonstrate dependence on the angle between the pump polarization and the crystal axes, thereby confirming their low symmetry, see Fig.10 b. As our detectors were not perfectly balanced the difference signal also contains fully symmetric coherent contributions. As mentioned above, one cannot compare the ratio of amplitudes for the doubly degenerate and fully symmetric phonons from the difference signal. However, from the comparison of isotropic and anisotropic signals we can estimate that $2c \leq a/10$ taking into account that the $E_g$ signal can be twice the maximum amplitude in optimal polarization geometry. To summarize, we experimentally show that the amplitude ratio roughly coincides with the Raman cross section ratio for both fully symmetric and doubly degenerate phonons separately, whereas of the ratios between fully symmetric and low symmetry phonons are drastically different for the time- and frequency domains.

Trying to explain the lack of low symmetry phonons for isotropic detection in the time-domain, we have to touch the topic related to the mechanism of coherent phonon generation. Based on the diverse phenomena, different generation mechanisms have been identified and theoretically formulated; see [7] and references therein. Among all the discussions, one topic over which has been debated for a long time is the nature of the generation mechanism for coherent phonons in opaque materials to which $Bi_2Te_3$ belongs. The question originated from the first experiments on semimetals and semiconductors, where large oscillations of fully symmetric modes were seen while other Raman modes were missing in the ultrafast pump-probe experiments. The results led to the theory of displacively excited coherent phonons, which declared that coherent phonons in opaque materials are generated by a different from Raman scattering process [34]. It was argued that coherent phonons are initiated because of a displaced quasi-equilibrium coordinate of the nuclear system set by the photoexcited carriers, and this mechanism only applies to modes which do not change the crystal symmetry. Later, after experimental observation of low-symmetry modes in semimetals [7], and the related theoretical development [35] it was suggested that the generation process in these semimetals is actually a particular case of coherent stimulated Raman scattering, which can generate modes of arbitrary, but even symmetry. Afterward, the theory of two stimulated Raman tensors [36] made the final attempt to combine impulsive excitations in transparent crystals with displacive excitations in absorbing materials under the framework of Raman scattering. It was pointed out that the difference in those two scenarios lies in distinct contributions to the driving force from virtual and real transitions. Subsequently, a model incorporating a finite lifetime of the driving forces was developed [37], and after a further refinement [38] used to explain the discrepancies between the time- and frequency- domain data in semimetals. Note, however, that according to measured resonance profiles of $E_g$ and $A_{1g}$ coherent phonons in Bi it was suggested that one can talk about the unified Raman mechanism only in the case when no distinction between hot luminescence and Raman process exists [39]. The data obtained in our study provide additional evidence for this hypothesis since the Raman data exhibiting $a^2 \approx c^2$ clearly disagree with the time-domain data suggesting $a >> c$. Nevertheless, for two fully symmetric modes in $Bi_2Te_3$ the unified Raman mechanism, including displacive (kinematic) excitation as a particular, resonance case, seems to work as their relative intensities in time- and frequency-domain measurements almost coincide. We would like to note here that the $Bi_2Te_3$ single crystal is an ideal, model system to address the mechanism of coherent phonon generation. Indeed, it has two fully symmetric and two doubly degenerate modes, which allow comparing the phonons of different symmetry and energy (in semimetals like Bi and Sb, we have only one fully symmetric mode, which frequency is larger than the frequency of doubly degenerate mode).

However, we cannot completely exclude the unified Raman mechanism for low symmetry modes at the current stage. Recall that the Raman intensity in a frequency-domain experiment is $I \propto \left| \sum_{kl} \vec{e}_s^k \chi_{kl} \vec{e}_i^l \right|^2$, where $\chi_{kl}$ is second rank Raman tensor and $\vec{e}_i$ and $\vec{e}_s$ are the polarization vectors for the incident and the scattered light. On the other hand, assuming that two tensor model [36] is correct, one can write $\frac{\Delta R}{R_o} \propto \sum_{kl} \vec{e}_s^k \pi_{kl} \chi_{kl} \vec{e}_i^l$, where $\pi_{kl}$ is a tensor related only to time-domain measurement, more exactly to coherent phonon generation [36,38]. There is a substantial difference between the $\pi_{kl}$ and $\chi_{kl}$ tensors because the decay $\Gamma$ appears twice in $\pi_{kl}$, once on in the dielectric function and a second time to account for the fact that intermediate states in absorbing materials have a finite lifetime. Instead, decay manifests itself only once in the $\chi_{kl}$ tensor, through the dielectric function [38]. The appearance of fully symmetric modes in the time-domain with the amplitude ratio comparable to the Raman ratio suggests $\pi_{kl}^{(A1g)} \approx \chi_k^{(A1g)}$. On the other hand, the lack of low symmetry modes for isotropic detection in the time-domain can occur only if $\pi_{kl}^{(Eg)} << \chi_k^{(Eg)}$. One possible reason for the realization of such a situation can be the lifetime of intermediate states for the generation of low symmetry phonons. Indeed, the driving force lifetime $1/\Gamma$ for fully symmetric phonons in Bi$_2$Te$_3$ defined by that of the photoexcited carriers is a few picoseconds, clearly satisfying the condition $\Omega \gg \Gamma$, where $\Omega$ is the phonon frequency. In this case, the initial coherent phonon amplitudes depend only weakly on the decay $\Gamma$. If the condition is reversed $\Gamma \gg \Omega$, which might happen for low symmetry modes, initial amplitudes decrease quite rapidly with larger $\Gamma$. If the real-space charge-density fluctuations serving as intermediate states for generation of low symmetry phonons have very short lifetimes satisfying $\Gamma \gg \Omega$ their amplitudes in the time-domain will be strongly suppressed. To check this hypothesis, temperature dependences of both time- and frequency domain measurements should be measured and compared.

V. Summary

To conclude, coherent and incoherent phonons in Bi$_2$Te$_3$ have been studied with ultrafast polarization-dependent laser and Raman scattering techniques. While in the frequency domain all Raman active phonons with comparable intensities were detected, in the time-domain only fully symmetric modes were observed with an isotropic detection scheme. To detect doubly degenerate modes anisotropic detection scheme must be used, which revealed that the ratios of fully symmetric to doubly degenerate amplitudes are different for the time- and frequency domains. Our data on the comparison between the amplitude of coherent phonons in time-domain pump-probe experiments and the cross section of spontaneous Raman scattering can help in elucidating long-standing questions about the coherent phonon generation, particularly the subtle differences between impulsive stimulated Raman scattering and the mechanism known as displacive excitation of coherent phonons [7, 34–39].

In some locations on the crystal surface additional phonon mode at 3.56 THz was observed and, based on its symmetry, revealed by polarization-dependent time-domain measurements, assigned to $A_1$ phonon of crystalline Te. This assignment is supported by the measurements of single crystal Te in the time- and frequency domains. Some of the observed features in this experimental study are not fully understood yet. For example, the negative, short-lived chirp of coherent phonons in stoichiometric $Bi_2Te_3$ is absent in the Te-rich regions, where both fully symmetric phonons of $Bi_2Te_3$ are also present. Next, the doubly degenerate $E_g$ modes with their large enough Raman polarizabilities did not appear in the frequency-domain data taken with isotropic detection, whereas for anisotropic detection their coherent amplitudes were shown to be significantly smaller than those of the fully symmetric $A_{1g}$ modes. More experimental and theoretical works are needed to understand these features. Nevertheless, the results presented here for coherent phonons in $Bi_2Te_3$ appear of general interest also for other three-dimensional topological insulators with a single Dirac cone.

ACKNOWLEDGMENTS

O.V.M. was in part supported by the Alexander von Humboldt Foundation in the framework of its alumni program, and the Russian Foundation for Basic Research through the grants 13-02-00263-a and 14-42-03576-p. J.F. and T.D. acknowledge the support by the Deutsche Forschungsgemeinschaft (DFG) through the SFB 767.

References

[1]  For a review, see, e.g. M. Z. Hasan and C. L. Kane, Rev. Mod. Phys. **82**, 3045 (2010).

[2]  F. Wilczek, Nature Physics **5**, 614 (2009).

[3]  K. S. Novoselov, A. K. Geim, S. V. Morozov, D. Jiang, M. I. Katsnelson, I. V. Grigorieva, S. V. Dubonos, and A. A. Firsov, Nature **438**, 197 (2005).

[4]  P. Hosur, Phys. Rev. **B 83**, 035309 (2011).

[5]  J. Maciejko, X. L. Qi, H. D. Drew, and S. C. Zhang, Phys. Rev. Lett. **105**, 166803 (2010).

[6]  J. A. Sobota, S.-L. Yang, D. Leuenberger, A. F. Kemper, J. G. Analytis, I. R. Fisher, P. S. Kirchmann, T. P. Devereaux, and Z.-X. Shen, Phys. Rev. Lett. **113**, 157401 (2014).

[7]  For a review, see, e.g. K. Ishioka, and O.V. Misochko, in Progress in Ultrafast Intense Laser Science V, Eds. K. Yamanouchi, A. Giullietti, and K. Ledingham, 23-64, Springer Series in Chemical Physics, Berlin (2010).

[8]  Y. L. Chen, J. G. Analytis, J.-H. Chu, Z. K. Liu, S.-K. Mo, X. L. Qi, H. J. Zhang, D. H. Lu, X. Dai, Z. Fang, S. C. Zhang, I. R. Fisher, Z. Hussain, and Z.-X. Shen, Science **325**, 178 (2009).

[9]  W. Richter, H. Kohler and C. R. Becker, Phys. Status Solidi (b) **84**, 619 (1977).

[10]  V. Wagner, G. Dolling, B. M. Powel, and G. Landweh, Phys. Status Solidi (b) **85**, 311 (1978).

[11]  W. Kullmann, J. Geurts, W. Richter, N. Lehner, H. Rauh, U. Steigenberger, G. Eichhorn, R. Geick, Phys. Status Solidi (b) **125,** 131 (1984).

[12]  A. Q. Wu, X. Xu, and R. Venkatasubramanian, Appl. Phys. Lett. **92**, 011108 (2008).

[13]  Y. Wang, X. Xu, and R. Venkatasubramanian, Appl. Phys. Lett. **93**, 113114 (2008).

[14]  Y. Wang, L. Guo, X. Xu, J. Pierce, and R. Venkatasubramanian, Phys. Rev. **B 88**, 064307 (2013).

[15] N. Kamaraju, S. Kumar, and A. K. Sood, EPL **92**, 47007 (2010).

[16] J. Flock, T. Dekorsy, and O.V. Misochko, Appl. Phys. Lett. **105**(1) 011902 (2014).

[17] A. Bartels, R. Cerna, C. Kistner, A. Thoma, F. Hudert, C. Janke, and T. Dekorsy, Rev. Sci. Instrum. **78**, 035107 (2007).

[18] R. Gebs, G. Klatt, C. Janke, T. Dekorsy, and A. Bartels, Opt. Express **18**, 5974 (2010).

[19] J. O. Jenkins, J. A. Rayne, and R. W. Ure, Jr., Phys. Rev. **B 5**, 3171 (1972).

[20] V. Chis, I. Yu. Sklyadneva, A. Kokh, V.A. Volodin, O. E. Tereshchenko, and E. V. Chulkov, Phys. Rev. **B 86**, 174304 (2012).

[21] O. V. Misochko, K. Ishioka, M. Hase, M. Kitajima, Journal of Physics: Condensed Matter **19**(15), 156227 (2007).

[22] O. V. Misochko, and M.V. Lebedev, Journal of Experimental and Theoretical Physics **120**(4), 651 (2015) [translated from ZhETF **147**, 750 (2015)].

[23] U. Fano, Phys. Rev. **72**, 26 (1947).

[24] O. V. Misochko, M.V. Lebedev, H. Schäfer, and T. Dekorsy, Journal of Physics: Condensed Matter **19**(40), 406220 (2007).

[25] K.G. Nakamura, J. Hu, K. Norimatsu, A. Goto, K. Igarashi, and T. Sasagawa, Solid State Communications **152**(10), 902 (2012).

[26] K. Norimatsu, J. Hu, A. Goto, K. Igarashi, T. Sasagawa, and K. G. Nakamura, Solid State Communications **157**, 58 (2013).

[27] S. Hunsche, K. Wienecke, T. Dekorsy, and H. Kurz, Phys. Rev. Lett. **75**, 1815 (1995).

[28] P. Tangney, and S. Fahy, Phys. Rev. **B 65**, 054302 (2002).

[29] O. V. Misochko, M.V. Lebedev, and T. Dekorsy, Journal of Physics: Condensed Matter **17**, 3015 (2005).

[30] V. Russo, A. Bailini, M. Zamboni, M. Passoni, C. Conti, C. S. Casari, A.L. Bassi, and C. E. Bottani, J. Raman Spectroscopy **39**, 205 (2008).

[31] Y. Li, V.A. Stoica, L. Endicott, G. Wang, C. Uher, and R. Clarke, Appl. Phys. Lett. **97**, 171908 (2010).

[32] R. He, Z. Wang, R. L. J. Qiu, C. Delaney, B. Beck, T. E. Kidd, C. C. Chancey, and X. P. A. Gao, Nanotechnology **23**, 455703 (2012).

[33] K. Norimatsu, M. Hada, S.Yamamoto, T. Sasagawa, M. Kitajima, Y. Kayanuma, and K. G. Nakamura, J. Appl. Phys. **117**, 143102 (2015).H. J. Zeiger, J. Vidal, T. K. Cheng, E. P. Ippen, G. Dresselhaus, and M. S. Dresselhaus, Phys. Rev. **B 45**, 768 (1992).

[35] G. A. Garrett, T. F. Albrecht, J. F. Whitaker, and R. Merlin, Phys. Rev.Lett. **77**, 3661 (1996).

[36] T. E. Stevens, J. Kuhl, and R. Merlin, Phys. Rev. **B 65**, 144304 (2002).

[37] D. M. Riffe, and A. J. Sabbah, Phys. Rev. **B 76**, 085207 (2007).

[38] J. J. Li, J. Chen, D. A. Reis, S. Fahy, and R. Merlin, Phys. Rev.Lett. **110**, 047401 (2013).

[39] A.A. Melnikov, O. V. Misochko, and S.V. Chekalin, Physics Letters **A 375**(19), 2017 (2011).

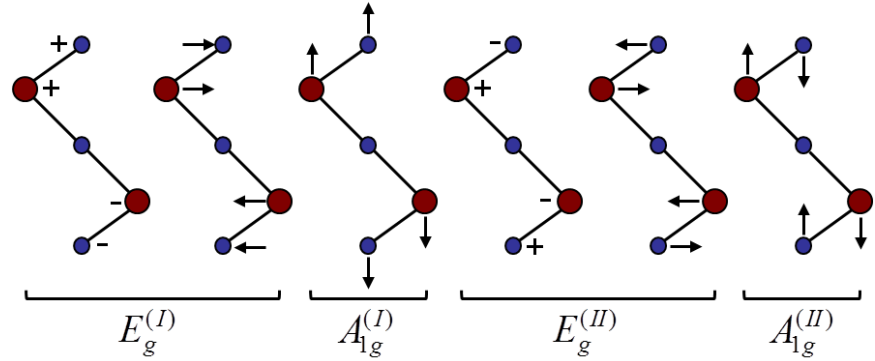

FIG. 1 (color online). The atomic displacements in Raman active modes of $Bi_2Te_3$ according to a factor group analysis. The small blue and the large red circles denote Te and Bi atoms, respectively.

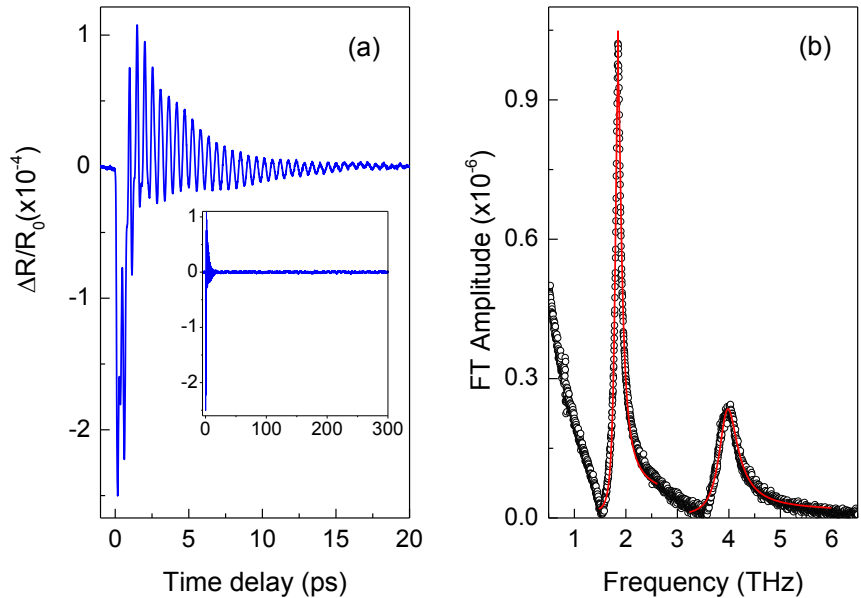

FIG. 2 (color online). A typical differential change in reflection in $Bi_2Te_3$ showing $A_{1g}$ coherent oscillations (a) and its fast FT spectrum (b). The inset in (a) shows the full trace measured out to long delay times. The open symbols in (b) are experimental points, while the solid red lines are the Fano fits ($q = 6.8$ and $\Gamma = 0.22$ for the high frequency mode; $q = 5.5$ and $\Gamma = 0.08$ for the low frequency mode).

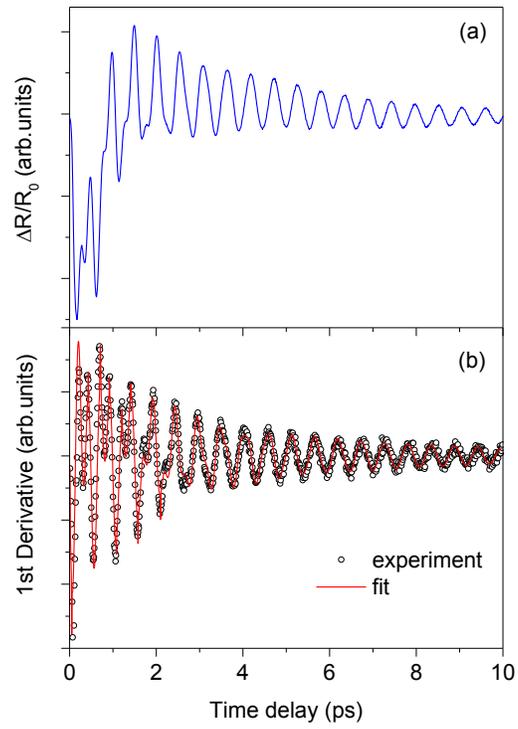

FIG. 3 (color online). A typical differential change in reflection in $Bi_2Te_3$ showing $A_{1g}$ coherent oscillations (a) and its derivative to separate the coherent part (b). In (b) circles are experimental points, whereas red solid line is the fit to two damped harmonic oscillators, see text.

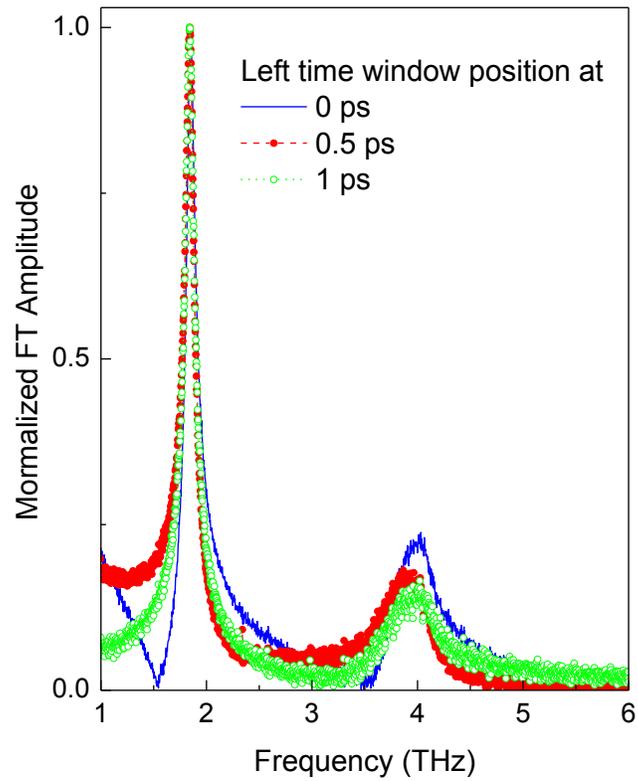

FIG. 4 (color online). FT spectra of the data in Fig.2a when left positions of the time-window for FT are shifted. Legends show different values of positions used. Data were normalized to the amplitude of the low frequency mode.

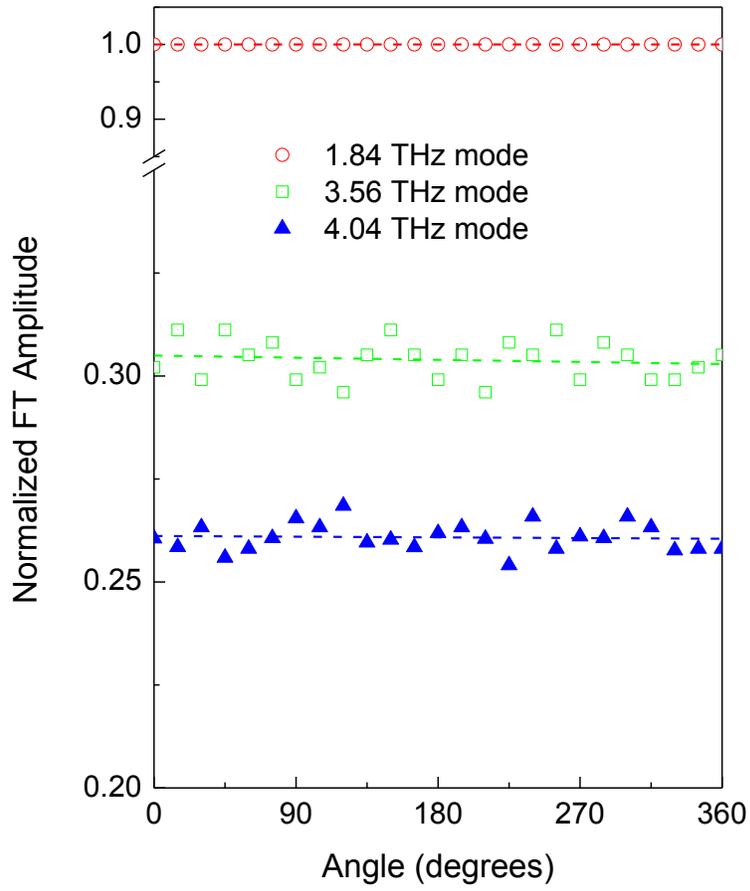

FIG. 5 (color online). Coherent phonon amplitudes of different modes obtained from the fast FT spectra as a function of angle. Legends specify the modes. The data were normalized to the amplitude of low-frequency $A_{1g}^{(I)}$ mode and the dashed lines are a guide to the eye.

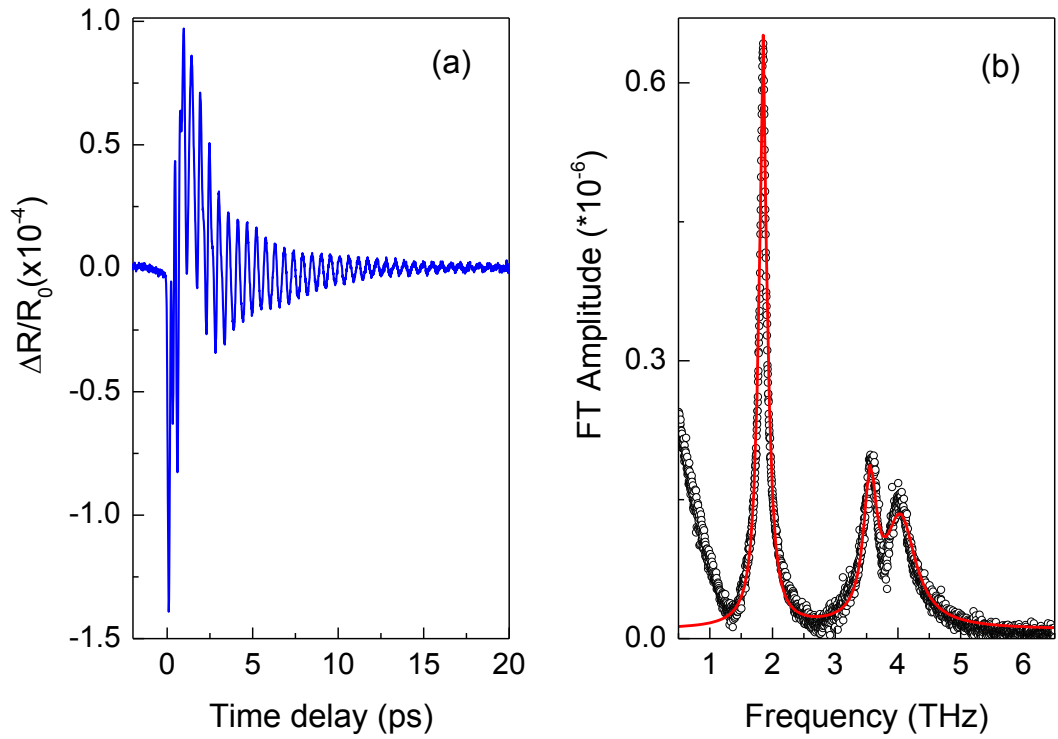

FIG. 6 (color online). Differential change in reflection in $Bi_2Te_3$ showing coherent oscillations (a) and its fast FT spectrum (b). The open symbols in (b) are experimental points, the solid line is a cumulative fit to 3 Lorentzian profiles.

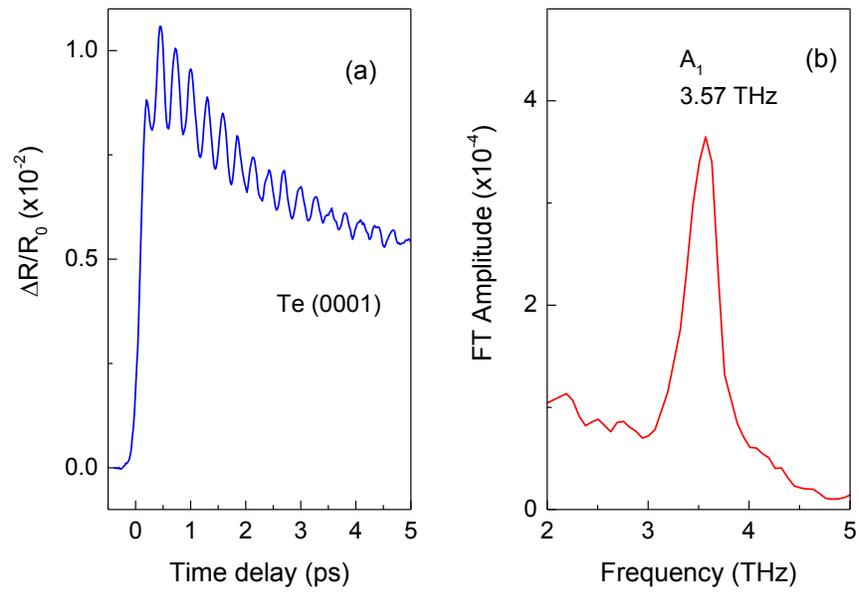

FIG. 7 (color online). Differential change in reflection for (0001) Te single crystal showing $A_1$ coherent oscillations (a) and its fast FT spectrum (b).

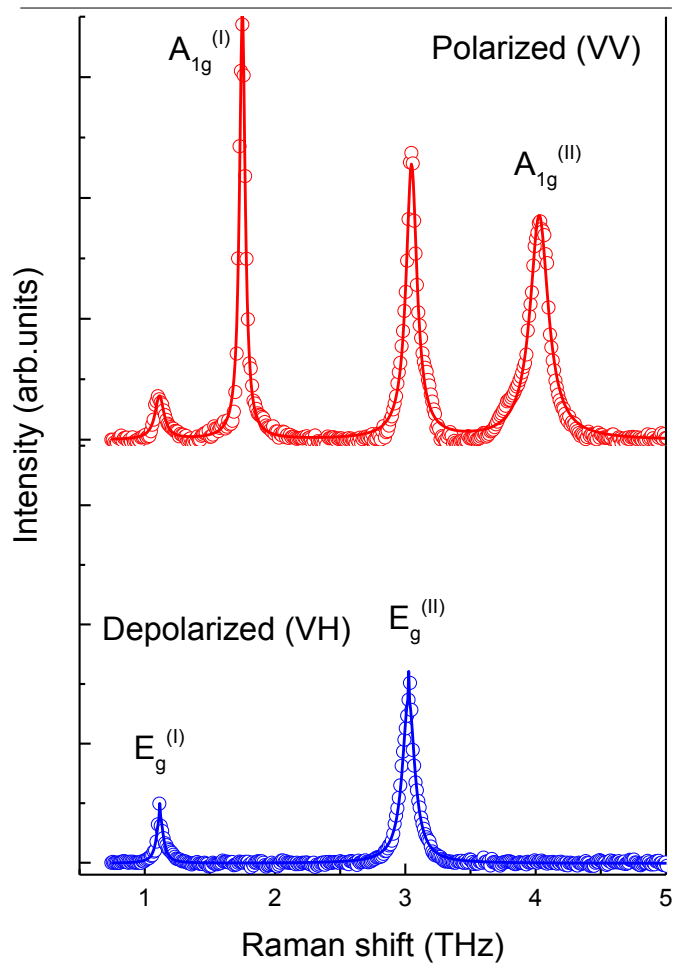

FIG. 8 (color online). Raman spectra of $Bi_2Te_3$ in backscattering geometry with an electric field parallel to the long dimension of the crystal (laser power at the sample position - 0.8 mW). The open symbols are experimental points, the solid lines are a cumulative fit to Lorentzian profiles. The upper trace is the polarized spectrum and the lower trace is the depolarized spectrum. Both the incident and scattered light polarizations belong to *xy*-plane. The spectra are vertically offset for clarity.

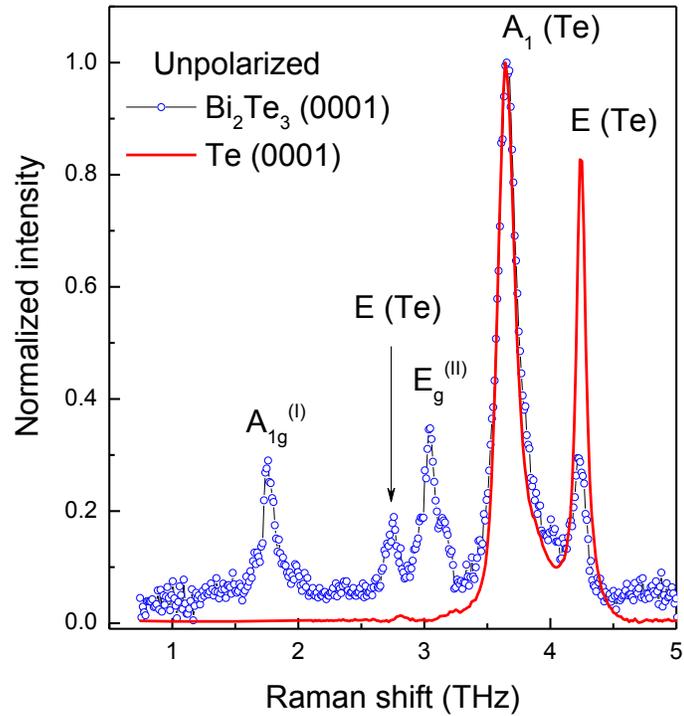

FIG. 9 (color online). Typical unpolarized Raman spectrum (open symbols) of $Bi_2Te_3$ in backscattering geometry without using analyzer showing additional lines (laser power at the sample position - 4 mW). For comparison purposes, unpolarized spectrum of (0001) Te is also shown by a red solid line (laser power at the sample position - 0.7 mW). Both spectra were normalized relative to the intensity of the strongest mode.

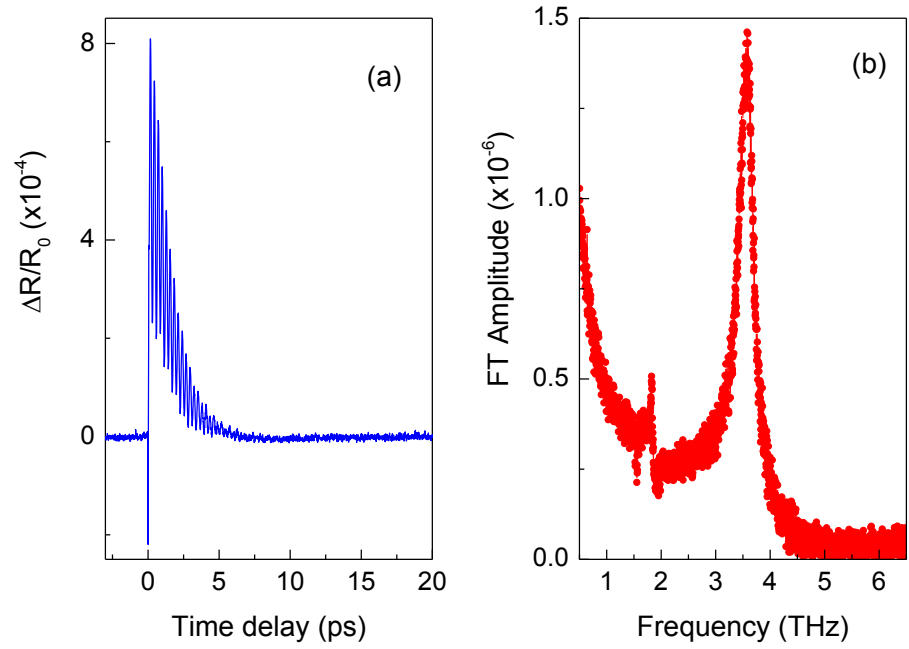

FIG. 10 (color online). Differential change in reflection in damaged $Bi_2Te_3$ showing coherent oscillations (a) and its fast FT spectrum (b).

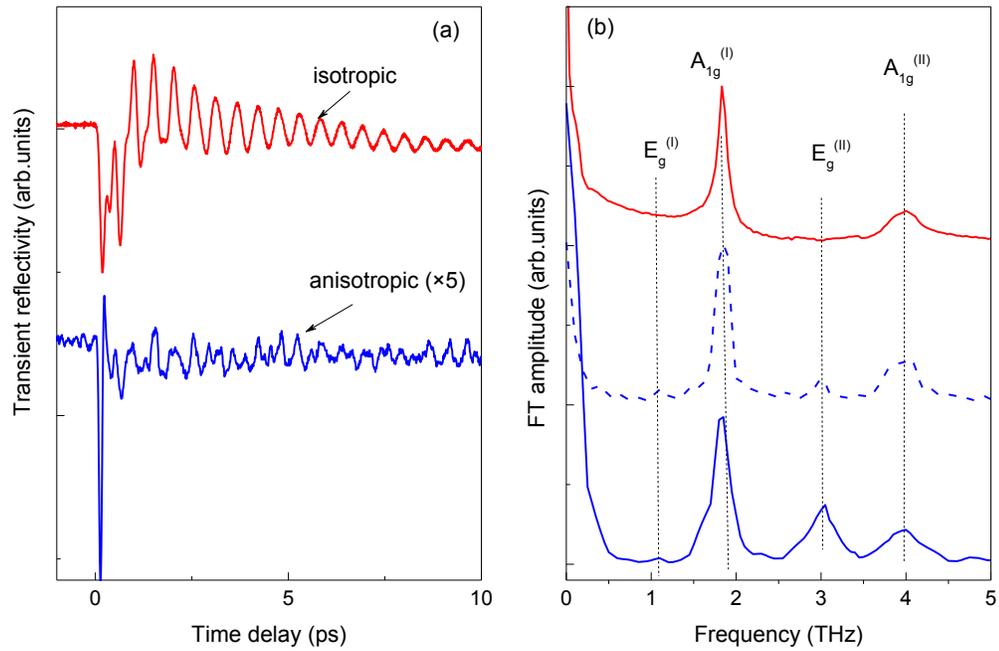

FIG. 11 (color online). Typical isotropic (upper trace) and anisotropic (lower trace) change in reflection for $Bi_2Te_3$ showing coherent oscillations (a) and their normalized fast FT spectra (b). The two lower spectra obtained with anisotropic detection differ from each other by the angle between the pump polarization and crystal axis, 0 (solid line) and $\pi/4$ (dashed line), respectively. The traces and spectra are vertically offset for clarity.

Table 1. Raman phonon parameters from the time- and frequency-domain measurements in Bi$_2$Te$_3$. All values are in THz.

| Mode | Time-domain | | Frequency-domain | |
| --- | --- | --- | --- | --- |
| | Frequency | Decay rate | Frequency | Decay rate |
| $A_{1g}^{(I)}$ | 1.84 | 0.2 | 1.80 | 0.21 |
| $A_{1g}^{(II)}$ | 4.04 | 0.8 | 4.03 | 0.67 |
| $E_g^{(I)}$ | 1.1 | - | 1.1 | 0.56 |
| $E_g^{(II)}$ | 3.02 | >0.7 | 3.05 | 0.47 |